\documentclass[10pt,a4paper,twoside]{article}
\usepackage{epsfig}
\usepackage{baltlat6}
\usepackage{array}
\usepackage{wrapfig}
\begin{document}
\ \
\vspace{0.5mm}
\setcounter{page}{143}
\vspace{8mm}

\titlehead{Baltic Astronomy, vol.\,17, 143--159, 2008}

\titleb{O-LIKE STARS IN THE DIRECTION OF THE NORTH AMERICA AND
PELICAN NEBULAE}

\begin{authorl}
\authorb{V. Strai\v{z}ys}{} and
\authorb{V. Laugalys}{}
\end{authorl}

\moveright-3.2mm
\vbox{
\begin{addressl}
\addressb{}{Institute of Theoretical Physics and Astronomy, Vilnius
University,\\  Go\v{s}tauto 12, Vilnius LT-01108, Lithuania;
straizys@itpa.lt, vygandas@itpa.lt}
\end{addressl}
}

\submitb{Received 2008 March 2; accepted 2008 April 10}

\begin{summary} In the area covering the complex of the North America
and Pelican nebulae we identified 13 faint stars with $J$--$H$ and
$H$--$K_s$ color indices which simulate heavily reddened O-type stars.
One of these stars is CP05-4 classified as O5\,V by Comer\'on and
Pasquali (2005).  Combining magnitudes of these stars in the passbands
$I_{\rm C}$, $J$, $H$, $K_s$ and [8.3] we were able to suspect that two
of them are carbon stars and five are late M-type AGB stars.
Interstellar extinction in the direction of these stars was estimated
from the background red clump giants in the $J$--$H$ vs.  $H$--$K_s$
diagram and from star counts in the $K_s$ passband.  Four or five stars
are found to have a considerable probability of being O-type stars,
contributing to the ionization of North America and Pelican.  If they
really are O-type stars, their interstellar extinction $A_V$ should be
from 16 to 35 mag.  Two of them seem to be responsible for bright E and
J radio rims discovered by Matthews \& Goss (1980).  \end{summary}


\resthead{O-like stars in the direction of North America and
Pelican}{V. Strai\v{z}ys, V. Laugalys}

\sectionb{1}{INTRODUCTION}

In the past there were numerous attempts to identify the star (or stars)
responsible for the ionization of the North America and Pelican nebulae
(the H\,II region W80).  The history of the search has been described
recently by Comer\'on \& Pasquali (2005).

The first true candidate for the ionizing star was HD\,199579 ($V$ =
6.0), a single-line spectroscopic binary of spectral class O6\,Ve,
proposed by Sharpless \& Osterbrock (1952).  The star is located in the
upper part of the North America Nebula and there are some signs of its
interaction with the nearby gas -- within about 0.5 degree from the star
the thermal radio flux in decimeter waves has somewhat larger intensity
(see, e.g, Matthews \& Goss 1980).  However HD\,199579 alone cannot be
responsible for the ionization of all the H\,II region since it is
located too far from the center of the complex.  Therefore a search for
other ionizing stars has been continued.  Herbig (1958) suggested that
the true ionizing star can be located behind the dust cloud L\,935 which
separates the North America and Pelican (hereafter NAP) nebulae.

Additional important information about the ionizing star was presented
by Bally \& Scoville (1980) who investigated the complex in $^{12}$CO
radio line at 2.60 mm.  From the analysis of line profiles they
concluded that the molecular complex expands with a velocity of 5 km/s
from the point near the peak of the thermal radio emission.  This
expansion was interpreted as the remnant of the ionization shock front
system from the H\,II region which was once formed by a young O-type
star (or stars) born off-center in the original molecular cloud about
3--8 million years ago.  When the shock reached the back edge of the
molecular cloud, an asymmetric flow of ionized gas was established,
depressurizing the inner H\,II region.  This outflow has formed a huge
outer H\,II region on the opposite side of the L\,935 dust and molecular
cloud, the present NAP nebulae.  In the direction of the Sun the front
edge of dust/molecular cloud was much thicker and remained almost
unaffected.  In an attempt to find the ionizing O-B stars behind the
L\,935 cloud Bally \& Scoville (1980) identified 11 infrared sources
which might be the candidates.

This model was extended by Wendker et al.  (1983) using the thermal
radio continuum observations at 11 cm.  The decimeter continuum reveals
a complicated picture of the flux distribution with a number of local
maxima, ridges, bright rims and other structural details.  To explain
these features, Wendker et al. proposed a model of the complex
containing a group of eight O-stars, ionizing so-called cavities in the
parent dust/molecular cloud.  They supposed that the resulting small
H\,II regions form the local radio flux maxima in the three ridges of
increased radio emission crossing the L\,935 cloud.  No candidates for
the ionizing stars were proposed.

One more feature which has played a significant role in the search for
the ionizing stars were the bright rim structures.  In the NAP
nebulae Pottasch (1956) described seven rims seen in the
optical.  The bright rims are usually located at the edges of dark
clouds where they meet the ionized region.  The rims are sharply
defined, especially from the dark side.  The brightest portion of each
rim is usually directed to the exciting star.  In space they have a form
of flat or curved sheets of ionized gas, and are best seen when viewed
edge-on.  Matthews \& Goss (1980) identified some of the Pottasch
optical rims in the thermal radio map at 49 cm and nine additional rims
which are seen only in radio -- optically they are completely obscured
by the dark cloud.  Four new radio rims were added by Wendker et al.
(1983).  The orientation of the rims was an important factor in
searching for places of the ionizing stars.

\sectionb{2}{THE CAMER\'ON AND PASQUALI STAR}

Comer\'on \& Pasquali (2005) in their search for the ionizing star have
used the $J$--$H$ vs.  $H$--$K_s$ and $K_s$ vs.  $H$--$K_s$ diagrams in
a circle of 0.5\degr\ radius centered on the coordinates RA (2000) =
20$^{\rm h}$55$^{\rm m}$17$^{\rm s}$, DEC (2000) =
+43\degr\,47\arcmin\,30\arcsec, near the geometric center of the
complex.  They identified 19 infrared objects lying near the
interstellar reddening line of O-type stars in the $J$--$H$ vs.
$H$--$K_s$ and $K_s$ vs.  $H$--$K_s$ diagrams.  For these objects the
infrared spectra between 1.5 and 2.4 $\mu$m were obtained, and two
early-type stars were found.  Finally, optical spectra of these two
stars were used for the identification of the star 2MASS
J205551.25+435224.6 (hereafter CP05-4) as the best candidate.  This star
was classified as O5\,V and recognized as the ionizing star for the
entire H\,II complex.  This star is present in the list of Bally \&
Scoville (1980) of the potential ionizing objects.

In one of our papers (Laugalys et al. 2006) we have measured the CP05-4
star in the 7-color {\it Vilnius} photometric system and confirmed that
it is a reliable candidate for a ionizing source of the nebulae.  If the
star indeed is of spectral type O5\,V, then {\it Vilnius} photometry
would give its $E_{Y-V}$ = 2.20 and $A_V$ = 9.2 for the normal
interstellar extinction law.  Adopting $V$ = 13.24 and $M_V$ = --4.5 we
obtain a distance of 520 pc, which is approximately the cloud distance.
If the star is more luminous, its distance should be larger.  Since we
have no accurate estimation of its luminosity, the relation of this star
to the ionization of the entire NAP complex is not secure.  If the star
is responsible for ionizing the complex, then, to avoid shielding of the
ultraviolet photons, it must be located well behind the L\,935 dust
cloud, in a relatively transparent space where the parent dust/molecular
cloud has been already destroyed.

Looking for young stellar objects (YSOs) in the NAP area, we have
analyzed stars measured in different photometric systems covering a
spectral range between 0.35 and 8.3 $\mu$m.
We have found some stars behind the L\,935 (Lynds 1962), or Tokyo 497
(Dobashi et al. 2005), dust/molecular cloud which are similar to heavily
reddened O-type stars located at a distance of the nebulae.  This
stimulated the present search for stars which can be contributors to the
ionization of the surrounding nebulae.  For the investigation a
3\degr\,$\times$\,3\degr\ area centered at J2000:  20$^{\rm
h}$\,56$^{\rm m}$, +44\degr\ was taken.

\sectionb{3}{MORE O-LIKE STARS BEHIND THE L\,935 CLOUD}

For the identification of potential ionizing stars behind the L\,935
cloud in the 2MASS catalog we applied a method similar to that used by
Comer\'on and Pasquali but with the following alterations:  (1) the
search area was extended along all the length of the L\,935 cloud; (2)
the magnitude $K_s$ limit was changed to $K_s < 5.6 + 1.8\,(H-K_s)$;
this limit includes all main-sequence O-type stars up to 550 pc and more
luminous stars at larger distances; (3) according to Strai\v{z}ys et al.
(2008), the ratio of color excesses $E_{J-H}/E_{H-K_s}$ in the
$Q$-parameter was taken 2.0, i.e., $Q_{JHK_s} = (J-H) - 2.0\,(H-K_s)$.
The limits set to $Q_{JHK_s}$ were the values between 0.05 and --0.15,
obtained from the intrinsic $Q_{JHK_s}$-value of O-type stars, --0.05,
taking into account the scatter of points in Figure 1 of paper
Strai\v{z}ys et al.  (2008).  One more restriction was put on the
accuracy of 2MASS photometry:  all objects with the errors in {\it J, H}
and $K_s$ magnitudes (given in the catalog) larger than 0.1 mag were not
considered.  According to Strai\v{z}ys \& Lazauskait\.e (2008), the
ratio $E_{J-H}/E_{H-K_s}$ depends slightly on the temperatures of stars:
with decreasing $T_{\rm eff}$ the ratio also decreases.  The expected
difference in slopes between O- and K-type stars is only about 3\,\%,
and this effect was neglected in the present study.

The objects, satisfying the above conditions, were checked in the Simbad
database, and all known stars of spectral class B and cooler were
rejected.  Among them quite numerous were N-type carbon stars.  A few
infrared objects, classified spectroscopically by Comer\'on \& Pasquali
(2005, Table 1) as carbon or AGB stars, were also excluded.  We also
excluded stars outside the L\,935 dust cloud and the Pelican Nebula
where the $A_V$ extinction lower than 10 mag is expected (Cambr\'esy et
al. 2002).  Most probably, these objects are not heavily reddened
O-stars.

The remaining 13 objects are listed in Table 1 which includes also
CP05-4, the confirmed O5\,V star.  The star No.\,12 was included in the
list despite its $Q_{JHK_s}$ = \hbox{--0.22;} see discussion in Section
7. All these objects hereafter will be called as `O-like stars'.  No
information about them is given in the Simbad database.  Their DSS2
magnitudes $V$, $F$ and $N$ were taken from the GSC-2.3 catalog
available at Simbad.  The $J$, $H$ and $K_s$ magnitudes were taken from
the 2MASS Point Source Catalog; in Table 1 their values are rounded to
two decimal places.  For most of the objects the MSX fluxes at 8.3
$\mu$m are available and are given in Table 1. The next columns give the
values of $Q_{JHK_s}$, maximum $A_V$ expected (see Section 6) and the
classification of the objects (see Sections 4, 5 and 7).

In Figure 1 the stars from Table 1 are plotted on a sky image,
together with the nebulae and the radio continuum at 21 cm intensity
isolines from the Canadian Galactic Plane Survey (Taylor et al. 2003).
Figure 2 shows the positions of these stars in the $K_s$ vs. $H$--$K_s$
diagram.


\def\k{\kern-5pt}
\begin{table}[!th]
\begin{center}
\vbox{\scriptsize\tabcolsep=2.3pt
\parbox[c]{110mm}{\baselineskip=9pt
{\smallbf Table 1.}{\small\ Stars in the North America and Pelican nebulae complex
simulating heavily reddened O-type stars.\lstrut}}
\begin{tabular}{rcccccccccccl}
\hline \hstrut
No. &  RA        &    DEC        &    $V$ &  $F$  &    $N$ &    $J$ &    $H$ &   $K_s$ & [8.3] & $Q_{JHK_s}$  &  $A_V$  &  Possible  \\
    & (J2000)    &  (J2000)      &    &       & (or $I$)  &        &        &         &  Jy     &        &  max.   &   type     \\		
\hline \hstrut
1~   & 20 49 57.23 & +44 22 53.8  &   --  &  --   &   --   &  12.01 &   9.31 &   7.98  &   --  &     0.04     &  23.4   &   O--B0?   \\
2~   & 20 50 35.05 & +44 26 29.6  &   --  &  --   &  16.96 &  \ 9.90 &  7.06 &   5.64  &  0.87 &     0.01     &  24.2   &  carbon?   \\
3~   & 20 52 46.79 & +44 02 30.8  &   --  &  --   &   --   &  11.99 &   9.21 &   7.85  &  0.15 &     0.06     &  20.5   &   AGB?     \\
4~   & 20 53 15.82 & +44 32 57.0  &   --  &  --   &   --   &  11.36 &   8.89 &   7.64  &   --  &   \k--0.03      &  18.7   &   O9--B0?  \\
5~   & 20 54 13.42 & +44 02 58.9  &   --  & 18.40 &  13.21 &  \ 8.37 &  6.48 &   5.58  &  0.78 &     0.08     &  21.1   &   AGB?     \\
6~   & 20 54 16.26 & +43 43 09.1  &   --  & 18.77 &  15.85 &  \ 8.24 &  6.36 &   5.46  &  1.06 &     0.08     &  22.0   &   AGB?     \\
7~   & 20 55 25.16 & +44 18 14.4  &  19.06 & 17.43 &  13.32 &  \ 7.96 &  6.20 &   5.29  &  0.86 &   \k--0.05      &  26.6   &   O5?     \\
8\rlap{*}~ & 20 55 51.25 & +43 52 24.6  &  13.24\rlap{*}& 10.69 &  \ 8.86 &   \ 6.36 &   5.51 &   5.04  &  0.84 &   \k--0.08      &  28.4   &   O5\,V    \\
9~   & 20 55 52.70 & +43 53 24.2  &    --  &  --   &  17.66 &  10.82 &   8.56 &   7.44  &  0.22 &     0.03     &  28.4   &   O9--B0?  \\
10~  & 20 57 36.47 & +44 04 55.9  &    --  &  --   &  17.77 &  10.63 &   8.57 &   7.48  &  0.22 &   \k--0.12      &  19.3   &   AGB?     \\
11~  & 20 58 06.73 & +43 55 14.1  &    --  &  --   &   --   &  12.88 &   8.66 &   6.57  &  0.80 &     0.04     &  26.6   &   O5, AGB? \\
12~  & 20 58 24.24 & +43 56 38.6  &    --  &  --   &   --   &  12.69 &   8.88 &   6.87  &  0.61 &   \k--0.22      &  26.6   &   AGB?     \\
13~  & 20 58 26.22 & +43 42 38.5  &  17.56 & 15.04 &  12.53 & \ 8.33 &   6.42 &   5.49  &  0.66 &     0.05     &  24.7   &  carbon?   \\
\tablerule
\noalign{\vskip1mm}
\multicolumn{13}{l}{\ \ {\smallbf Note:}\small\ No.\,8 = CP05-4, its $V$ magnitude is from
Laugalys et al. (2006).}\\
\end{tabular}
}
\end{center}
\vspace{-2mm}
\end{table}


The `O-like stars' can be either O-type stars or high-luminosity stars
of other spectral classes located behind the L\,935 cloud, i.e. farther
than 500--600 pc.  All cooler main-sequence stars (including B-stars)
behind the cloud are eliminated since in the $K_s$ passband they are
fainter than O-stars.  Normal K and M red giants are excluded by the
$Q_{JHK_s}$\,$<$\,0.15 criterion.  However, B-stars of higher
luminosities (from IV to I), supergiants of spectral classes A--F5 and
AGB stars of the latest spectral classes (M6--M10 for oxygen-rich giants
and N-stars for carbon-rich giants) can intervene the $J$--$H$ vs.
$H$--$K_s$ and $K_s$ vs.  $H$--$K_s$ diagram regions used to isolate
possible O-type stars.


\begin{figure}[!th]
\vbox{
\centerline{\psfig{figure=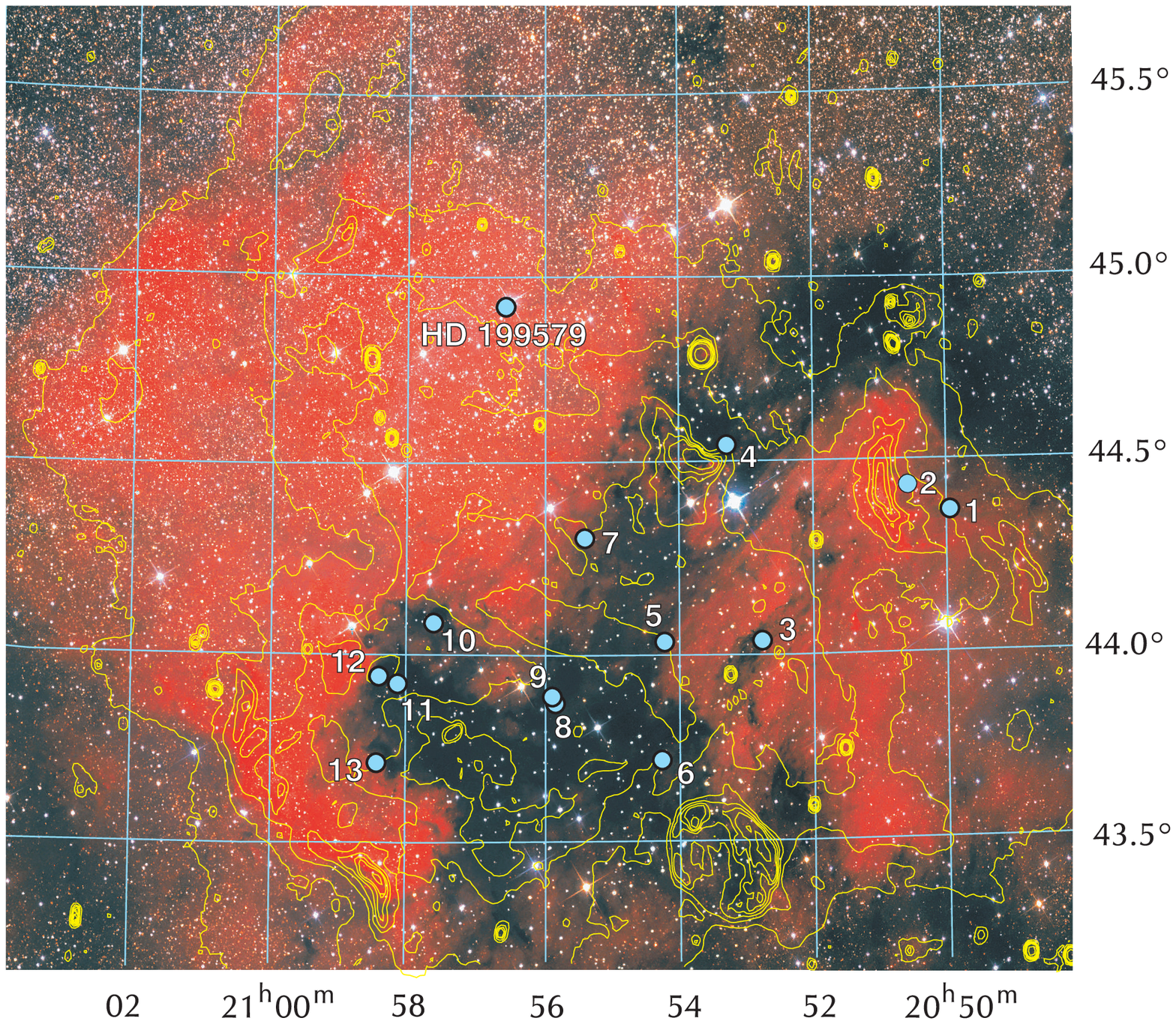,width=125mm,angle=0,clip=}}
\vspace{1mm}
\captionb{1}{The map of the NAP nebulae region with the radio continuum
isolines from the Canadian Galactic Plane Survey and the stars from
Table 1. The radio isolines correspond to the $T_{\rm b}$ values of 6,
8, 10, 12, 14, 16, 18, 20, 25 and 50 K. The oval feature of
15\arcmin\,$\times$\,20\arcmin\ size at 20$^{\rm h}$53.3$^{\rm m}$,
+43\degr\,27\arcmin\ is the supernova remnant SNR 084.2-00.8 located at
a distance of 4.5 kpc (Matthews et al. 1977; Kaplan et al. 2004).  Most
of other small radio sources in the area are extragalactic objects
(Matthews \& Goss 1980).}
}
\end{figure}


\begin{figure}[!th]
\vbox{
\centerline{\psfig{figure=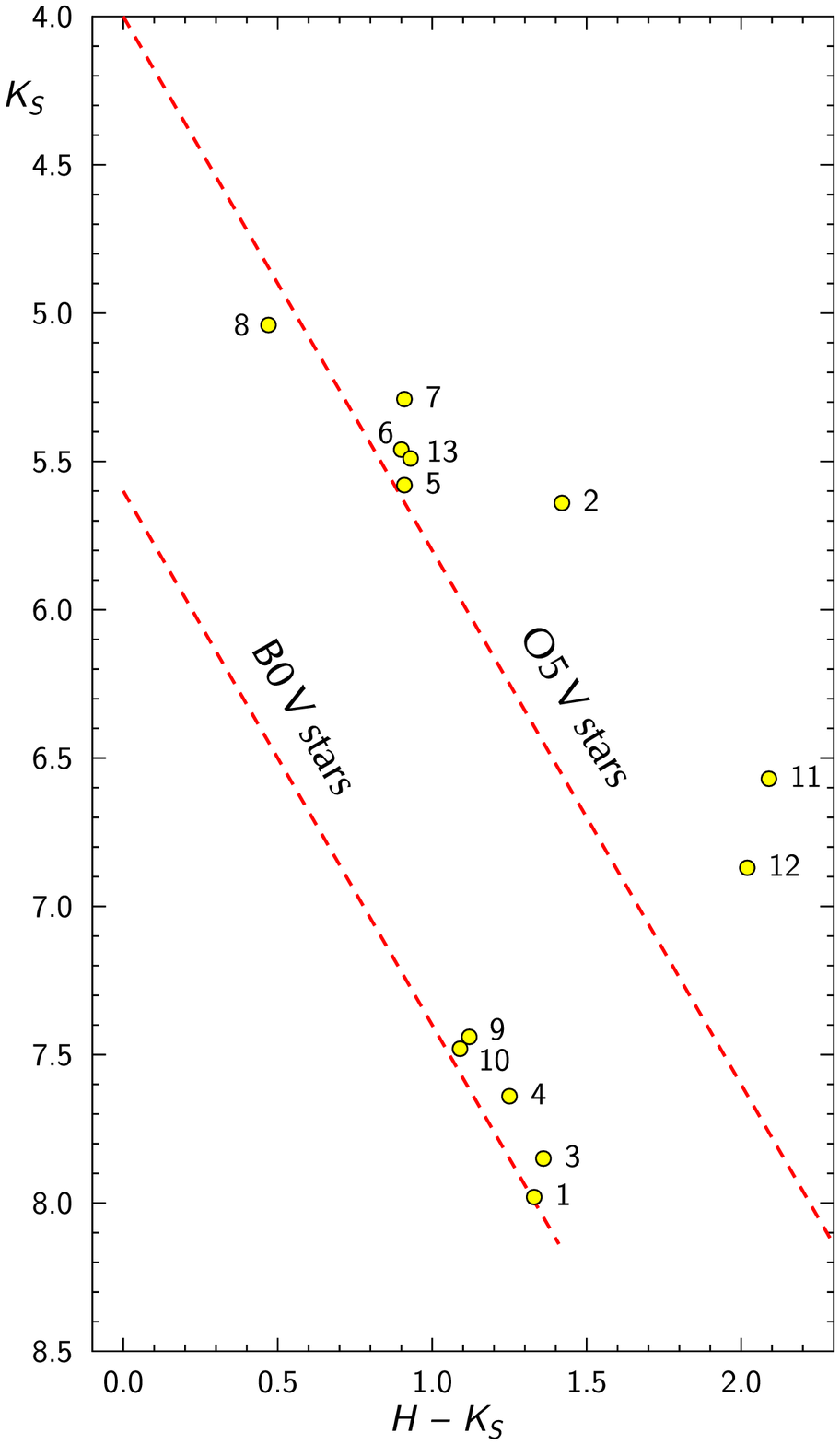,height=146mm,angle=0,clip=}}
\vspace{-.5mm}
\captionb{2}{Color-magnitude diagram $K_s$ vs. $H$--$K_s$ for the stars from
Table 1. The two parallel lines are the interstellar reddening lines for
the stars of spectral types O5\,V and B0\,V at a distance of 550 pc.}
}
\end{figure}



\begin{figure}[!th]
\vbox{
\centerline{\psfig{figure=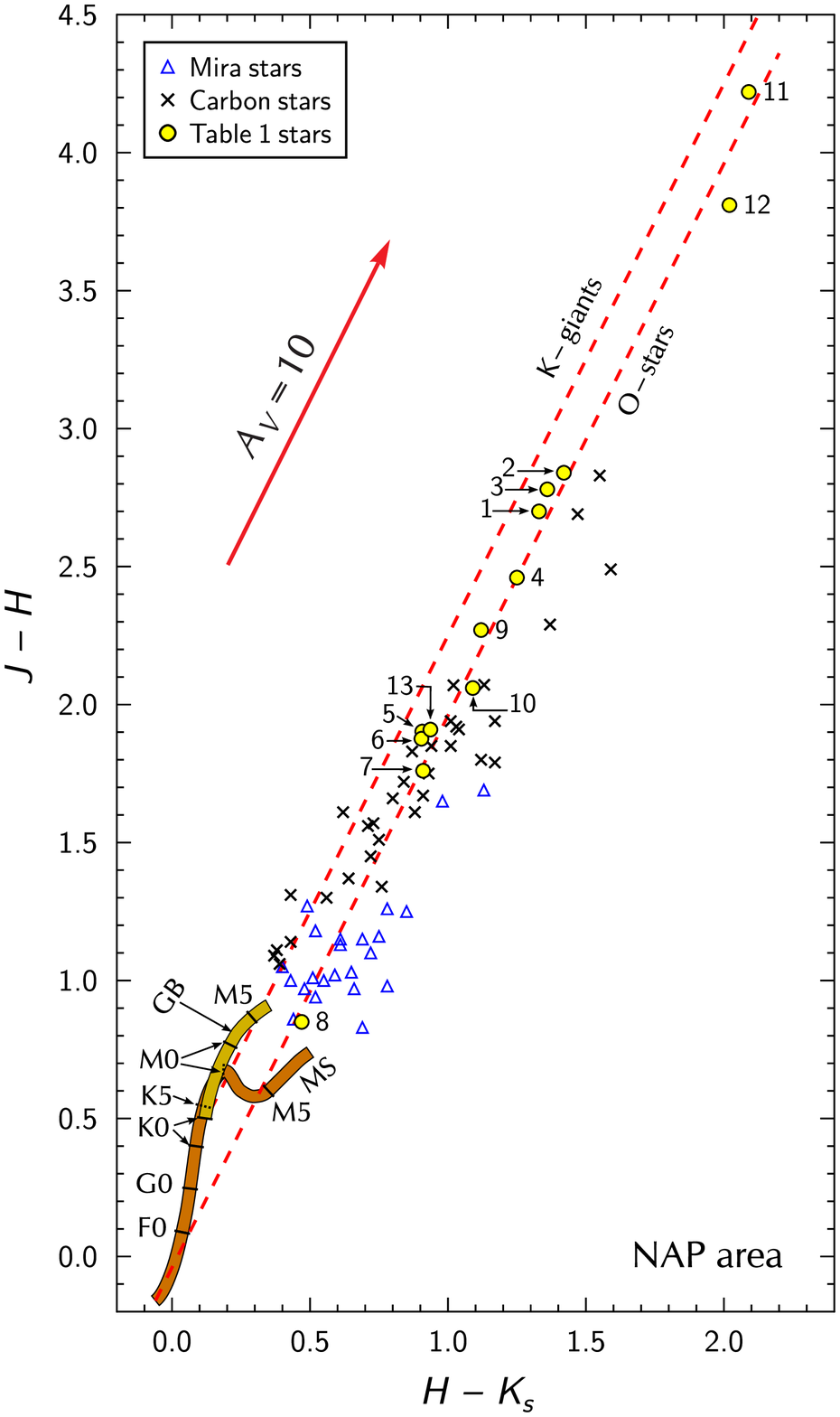,width=100mm,angle=0,clip=}}
\vspace{-.5mm}
\captionb{3}{The $J$--$H$ vs.  $H$--$K_s$ diagram with the intrinsic main
sequence (MS) and giant branch (GB).  Two parallel broken lines (in red)
are the interstellar reddening lines of O-type stars and red clump
giants.  Crosses designate known N-type carbon stars and blue triangles
O-rich Mira variables located in the NAP area.
The numbered yellow circles indicate the stars from Table 1.}
}
\end{figure}


\begin{figure}[!th]
\vbox{
\centerline{\psfig{figure=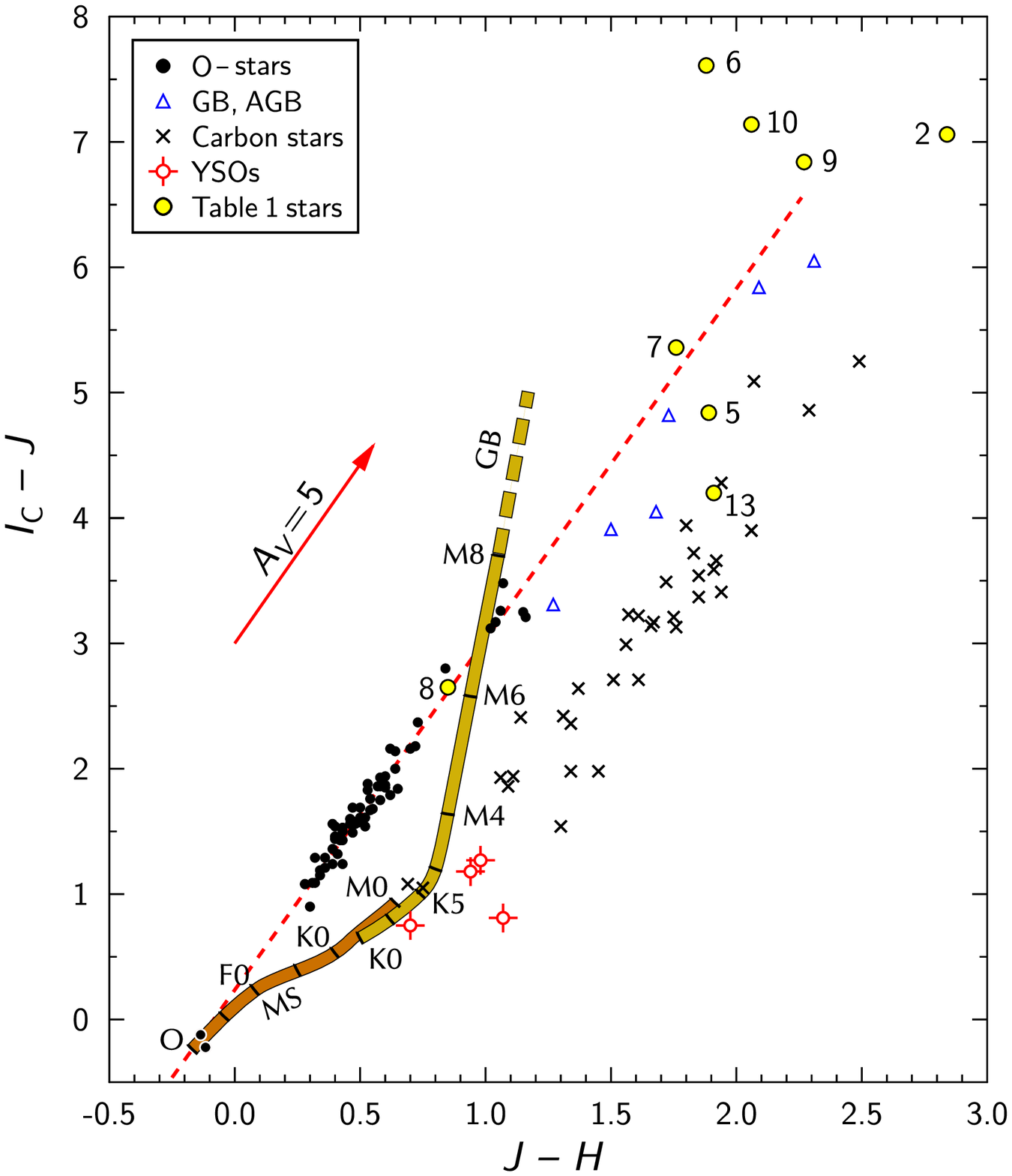,width=125mm,angle=0,clip=}}
\vspace{-.5mm}
\captionb{4}{The $I_{\rm C}$--$J$ vs.  $J$--$H$ diagram with the intrinsic
main sequence (MS) and red giant branch (GB) (thick orange and
yellow lines).
Dots are O--B1 stars in the NAP and Cyg OB2 association areas, defining
the interstellar reddening line.  Crosses designate known N-type
carbon stars in the NAP area and blue triangles designate GB and AGB
stars classified by Comer\'on \& Pasquali (2005).  The numbered yellow
circles indicate the stars from Table 1.}
}
\vskip20mm
\end{figure}

Figure 3 shows the $J$--$H$ vs.  $H$--$K_s$ diagram with the intrinsic
lines of main sequence and late-type giants and the two reddening lines
corresponding to red clump giants (G8--K2\,III, $Q_{JHK_s}$ = 0.22) and
O--B stars ($Q_{JHK_s}$ = --0.05).  The crosses represent known N-type
carbon stars (Table 2) selected in the NAP area of
3\degr\,$\times$\,3\degr\ size with the center given at the end of
Section 2. The blue triangles are known Mira-type variables (Table 3)
selected in a larger area around the NAP nebulae, with RA between
$20^{\rm h}30{\rm m}$ and $21^{\rm h}30{\rm m}$ and DEC from +41\degr\
to +46\degr.  Most (if not all) Mira variables are oxygen-rich (M-type)
giants.  It is evident that the reddened carbon stars and Miras
completely cover the reddening line of O-type stars.  Thus, the most
important task is to identify these types of objects.


\begin{table}[!th]
\begin{center}
\vbox{\small\tabcolsep=4pt
\parbox[c]{110mm}{\baselineskip=9pt
{\smallbf Table 2.}{\small\ Color indices of carbon stars in the 3\degr\,$\times$\,3\degr\
NAP area (rounded to two decimal places).\lstrut}}
\begin{tabular}{cccc|@{\huad}cccc}
\hline \hstrut
CGCS    &  $N$--$J$ &  $J$--$H$  &  $H$--$K_s$ &
CGCS    &  $N$--$J$ &  $J$--$H$  &  $H$--$K_s$  \\
\tablerule
4982    &  4.28 &  1.94 &  1.01  &  5049    &  3.14 &  1.66 &  0.80 \\ [-2pt]
4986    &  1.93 &  1.06 &  0.39  &  5058    &  3.23 &  1.57 &  0.73 \\ [-2pt]
4992    &  1.05 &  0.75 &  0.17  &  5061    &  1.86 &  1.09 &  0.37 \\ [-2pt]
4993    &  5.09 &  2.07 &  1.02  &  5071    &  3.94 &  1.80 &  1.12 \\ [-2pt]
4996    &  3.59 &  1.91 &  1.04  &  5075    &  2.99 &  1.56 &  0.71 \\ [-2pt]
4998    &  3.49 &  1.72 &  0.84  &  5084    &  2.64 &  1.37 &  0.64 \\ [-2pt]
5002    &  3.90 &  2.06 &  1.11  &  5085    &  1.81 &  1.79 &  1.17 \\ [-2pt]
5003    &  3.13 &  1.76 &  0.91  &  5091    &  1.94 &  1.11 &  0.38 \\ [-2pt]
5007    &  3.66 &  1.92 &  1.03  &  5095    &  3.17 &  1.67 &  0.91 \\ [-2pt]
5010    &  3.54 &  1.85 &  0.94  &  5099    &  2.41 &  1.14 &  0.43 \\ [-2pt]
5014    &  3.41 &  1.94 &  1.17  &  5105    &  1.98 &  1.45 &  0.72 \\ [-2pt]
5017    &  2.71 &  1.61 &  0.88  &  5110    &  1.98 &  1.34 &  0.76 \\ [-2pt]
5020    &  3.72 &  1.83 &  0.87  &  5124    &  2.71 &  1.51 &  0.75 \\ [-2pt]
5021    &  3.37 &  1.85 &  1.01  &  5135    &  1.54 &  1.30 &  0.56 \\ [-2pt]
5027    &  1.08 &  0.69 &  0.22  &  CP05-10 &  4.86 &  2.29 &  1.37 \\ [-2pt]
6876\rlap{*}   &  3.21 &  1.75 &  0.93  &  CP05-16 &  --   &  2.69 &  1.47 \\ [-2pt]
5033    &  2.42 &  1.31 &  0.43  &  CP05-18 &  5.25 &  2.49 &  1.59 \\ [-2pt]
5035    &  3.22 &  1.61 &  0.62  &  CP05-19 &  --   &  2.83 &  1.55 \\
\tablerule
\noalign{\vskip1mm}
\multicolumn{8}{l}{\ \ {\smallbf Note:}\small\ CGCS 6876 = CP05-6.}\\
\end{tabular}
}
\end{center}
\vskip-2mm
\end{table}

\vskip-2mm


\begin{table}[!th]
\begin{center}
\vbox{\small\tabcolsep=4pt
\parbox[c]{110mm}{\baselineskip=9pt
{\smallbf Table 3.}{\small\ Color indices of Mira variables in the NAP area (rounded to
two decimal places).\lstrut}}
\begin{tabular}{lcc|lcc}
\hline \hstrut
Name  & $J$--$H$  &  $H$--$K_s$  &  Name & $J$--$H$  &  $H$--$K_s$ \\
\tablerule
BH Cyg &  0.97 &   0.48   &      V603 Cyg  &    1.69 &   1.13  \\ [-2pt]
BL Cyg &  0.83 &   0.69   &      V607 Cyg  &    0.94 &   0.52  \\ [-2pt]
DG Cyg &  1.27 &   0.49   &      V780 Cyg  &    1.15 &   0.69  \\ [-2pt]
V506 Cyg  &  1.26 &   0.78   &   V1223 Cyg &    1.16 &   0.75  \\ [-2pt]
V528 Cyg  &  1.25 &   0.85   &   V1225 Cyg &    1.00 &   0.43  \\ [-2pt]
V580 Cyg  &  1.03 &   0.65   &   V1232 Cyg &    1.00 &   0.55  \\ [-2pt]
V584 Cyg  &  1.02 &   0.59   &   V1234 Cyg &    1.13 &   0.61  \\ [-2pt]
V593 Cyg   &  0.97 &   0.66   &  V1242 Cyg &    1.18 &   0.52  \\ [-2pt]
V596 Cyg   &  0.86 &   0.44   &  V1243 Cyg &    0.98 &   0.78  \\ [-2pt]
V597 Cyg   &  1.15 &   0.61   &  V1480 Cyg &    1.05 &   0.40  \\ [-2pt]
V600 Cyg   &  1.10 &   0.72   &  V1660 Cyg &    1.65 &   0.98  \\ [-2pt]
V601 Cyg  &  1.01 &   0.51   &             &         &         \\
\tablerule
\end{tabular}
}
\end{center}
\end{table}


\sectionb{4}{THE I--J vs. J--H DIAGRAM}

For the identification of carbon stars the $I_{\rm C}$--$J$ vs.
$J$--$H$ diagram, shown in Figure 4, can be used.  Here $I_{\rm C}$ is
the far-red magnitude close to the Cousins system.  The intrinsic lines
of normal main-sequence stars and cool giants were calculated in the
following way.  The intrinsic color indices $I_{\rm C}$--$J$ for
main-sequence stars and K-giants were calculated from the tabulation of
$V$--$J$ by Koornneef (1983) and $V$--$I_{\rm C}$ by Strai\v{z}ys
(1992).  The same color indices for M0--M6 giants were calculated from
$V$--$J$ by Koornneef (1983) and $V$--$I_{\rm C}$ by The et al.  (1990).
The intrinsic color indices $J$--$H$ for main-sequence stars and
late-type giants were taken from Bessell \& Brett (1988) after their
transformation to the 2MASS system by the equations given in Carpenter
(2001).

The intrinsic line of M giants was extended up to the spectral type
M8\,III taking its DENIS-based $I$--$J$ = 3.8 and $J$--$H$ = 1.1 from
Glass \& Schultheis (2002) and Groenewegen \& Blommaert (2005).  The
difference between the 2MASS and DENIS systems was neglected.  This
value of $J$--$H$ is confirmed by 2MASS results for the coolest
Mira-type variables:  R Cas (M6e--M10, $P$ = 430 d), W And (M7-Se,
S6,1e--S9,2e, $P$ = 397 d) and RU Her (M6e--M9, $P$ = 485 d).  Their
$J$--$H$ values from 2MASS are 1.01, 1.09 and 1.01, respectively. The
intrinsic line of M-giants runs almost vertically due to increasing
absorption in the TiO band at 850 nm.

To define the interstellar reddening line, in Figure 4 we plotted 58
O-type stars belonging to the Cyg OB2 association and listed in Table 1
of Strai\v{z}ys et al.  (2008).  Color indices $I_{\rm C}$--$J$ and
$J$--$H$ of these stars were calculated either from the $I_{\rm C}$
magnitudes given in Droege et al.  (2006) or from the DSS2 photographic
$N$ magnitudes (given in the GSC-2.3 catalog, Simbad), and the $J$ and
$H$ magnitudes from 2MASS.\footnote {~We assumed that the far-red
magnitude systems of $I_{\rm C}$ and $N$ coincide.  This assumption may
not be strictly correct as the response curves of both magnitude systems
and their mean wavelengths are slightly different (806 nm for $I_{\rm
C}$ and 840 nm for $N$, see the {\it Asiago Database on Photometric
Systems}, Fiorucci \& Munari (2003).  However, the positions of reddened
O-stars do not differ systematically when using either $I_{\rm C}$ from
Droege et al.  (2006) or $N$ from GSC-2.3.} Additionally, we plotted two
O-type stars with small interstellar reddening (S Mon and 10 Lac) and
the O5-type star CP05-4 located in the background of the L\,935 dust
cloud (Comer\'on \& Pasquali 2005).  The slope of the reddening line is
$E_{I_{\rm C}-J}$ / $E_{J-H}$ = 2.84 which is considerably larger than
the value 1.78 calculated from the $A_{\lambda}$ / $E_{B-V}$ ratios
given by Fitzpatrick (1999, Table 2).  The reason for this disagreement
probably is related to deviations of the Droege et al.  (2006) and the
DSS2 far-red systems from the standard $I_{\rm C}$ system, which itself
has a poor definition and a number of versions.

Due to the band-width effect the ratio $E_{I_{\rm C}-J}$ / $E_{J-H}$
depends slightly on the temperature of stars (Strai\v{z}ys \&
Lazauskait\.e 2008):  for K-type stars the ratio is by 3\,\% {\it
larger} than for O-type stars.  However, taking into account a very poor
knowledge of response functions of the far-red passbands ($I$ or $N$),
this effect in Figure 4 is too small to be significant.

In Figure 4 we also plotted known N-type carbon stars selected in the
NAP 3\degr\,$\times$\,3\degr\ area (Table 2).  Their far-red $I_{\rm C}$
magnitudes were taken mostly from the GSC-2.3 (i.e., DSS2 $N$
magnitudes).  For three stars we used the magnitudes from Droege et al.
(2006) and for one star (CGCS\,4996) from the {\it INT Photometric
H$\alpha$ Survey of the Northern Galactic Plane}, IPHAS (Drew et al.
2005; Gonz\'alez-Solares et al. 2008).  The response curve of the IPHAS
far-red magnitude $i$ is shifted blueward from $I_{\rm C}$, and has a
mean wavelength of 774 nm.  The $i$ magnitude was transformed from the
IPHAS system by the equation $N$ = $i$ -- 1.10 obtained from a
comparison of six other carbon stars with both $N$ and $i$ magnitudes
available.

It is evident that the reddening line of carbon stars is more or less
parallel to the reddening line of O--B1 type stars but lies about 1.4
mag lower.  This makes possible to identify carbon stars even at the
level of relatively low accuracy of the photographic DSS $N$ magnitudes.
A considerable scatter of carbon stars around their mean line may be the
result of low accuracy of their photographic far-red magnitudes,
possible variability and the intrinsic peculiarities in their spectral
energy distributions.

However, the $I_{\rm C}$--$J$ vs.  $J$--$H$ diagram is almost useless
for the separation of O-type stars from oxygen-rich giants of spectral
types M6\,III and cooler.  In Figure 4 we plot three GB and three AGB
stars (triangles) classified by Comer\'on \& Pasquali (2005), their
$N$ magnitudes being taken from GSC-2.3.  All of them lie between
the O-type reddening line and the carbon star sequence.  Cooler AGB
stars with the TiO-dominated spectra (Mira variables) in Figure 4 should
cover both the reddening line of O-stars and the region above it.

To exhibit the location of YSOs, we plotted in Figure 4 two known T
Tauri type stars and two H$\alpha$ emission stars which lie in the Gulf
of Mexico (red crossed circles).  With increasing reddening, these stars
can overlap carbon-rich stars but will not be mixed with O-type stars.

The O-like stars in Figure 4 are plotted as yellow circles, with their
numbers from Table 1. Their far-red $N$ magnitudes (when available) are
taken from GSC-2.3.  Of these stars only CP05-4 has another source of
the far-red magnitude:  $I_{\rm C}$ = 9.00 in Droege et al.  (2006).
Other three stars (3, 7 and 13) were observed in the IPHAS project
(Gonz\'alez-Solares et al. 2008).  For the sake of uniformity, we have
opted to use the $N$ magnitudes for these stars, too.

Figure 4 shows that only two stars from Table 1, Nos. 7 and 9, lie
almost on the extension of the reddening line of O-stars.  They are the
best candidates for O-type stars.  Probably, stars Nos. 2 and 13 are
carbon stars, while Nos. 5, 6 and 10 are oxygen-rich AGB stars (M-type
Miras or other M-type long-period variables).  Although their belonging
to O-type stars is doubtful, they should remain in the list for future
verification of spectral types by spectroscopic means.
Photometric classification of stars 1, 3, 4, 11 and 12 is impossible at
this moment since they have no measured far-red magnitudes.  They remain
in the list of suspected O-type stars.

\sectionb{5}{SPECTRAL ENERGY DISTRIBUTIONS}

With the aim to find further evidence that at least some of the selected
stars belong to early-type stars, we calculated their spectral energy
distributions (SEDs) between 0.8 and 8.3 $\mu$m using the data given in
Table 1. The values of $\log\,\lambda\,F_{\lambda}$ for the $N$ (or
$I_{\rm C}$), $J$, $H$ and $K_s$ magnitudes and the MSX 8.3 $\mu$m flux
were obtained as described in Strai\v{z}ys \& Laugalys (2007, p.
341--342).

The SED curves obtained are shown in Figure 5:  probable O-type stars in
panel (a) and probable AGB stars in panel (b).  The star numbers in the
inserts correspond to Table 1. All probable O-stars have their SED
maxima at 2.2 $\mu$m ($K$ magnitude), except of the CP05-4 star which
has maximum at $H$ magnitude.  The reddest is No.\,4 but it has not been
measured in the MSX band at 8.3 $\mu$m.  The form of the SED curves
suggests that the `O-like' stars have no infrared excesses at $>$\,2.2
$\mu$m, i.e., they do not belong to YSOs or AGB stars with dust
envelopes.

SEDs of the suspected AGB stars (Figure 5, panel b) are not very
different from those given in panel (a).  At such high interstellar
reddenings the intrinsic differences of SEDs between O-type stars and
late M-type stars without circumstellar dust envelopes in the 1--3
$\mu$m range become negligible if the photometric passbands do not
contain strong TiO or H$_2$O bands.  The CO bands longward of 1.56 and
2.32 $\mu$m, C$_2$ bands longward of 1.77 $\mu$m and H$_2$O bands at
1.3--1.5 and 1.7--2.0 $\mu$m, present in the spectra of AGB stars, are
too faint to create a measurable photometric effect in the broad-band
$J$--$H$ and $H$--$K_s$ color indices.  However, these spectral features
are easily observable in infrared spectra at a resolution of
$\lambda$/$\Delta\lambda$ = 240 (Comer\'on \& Pasquali 2005).


\begin{figure}[!th]
\vbox{
\parbox[t]{61mm}{\psfig{figure=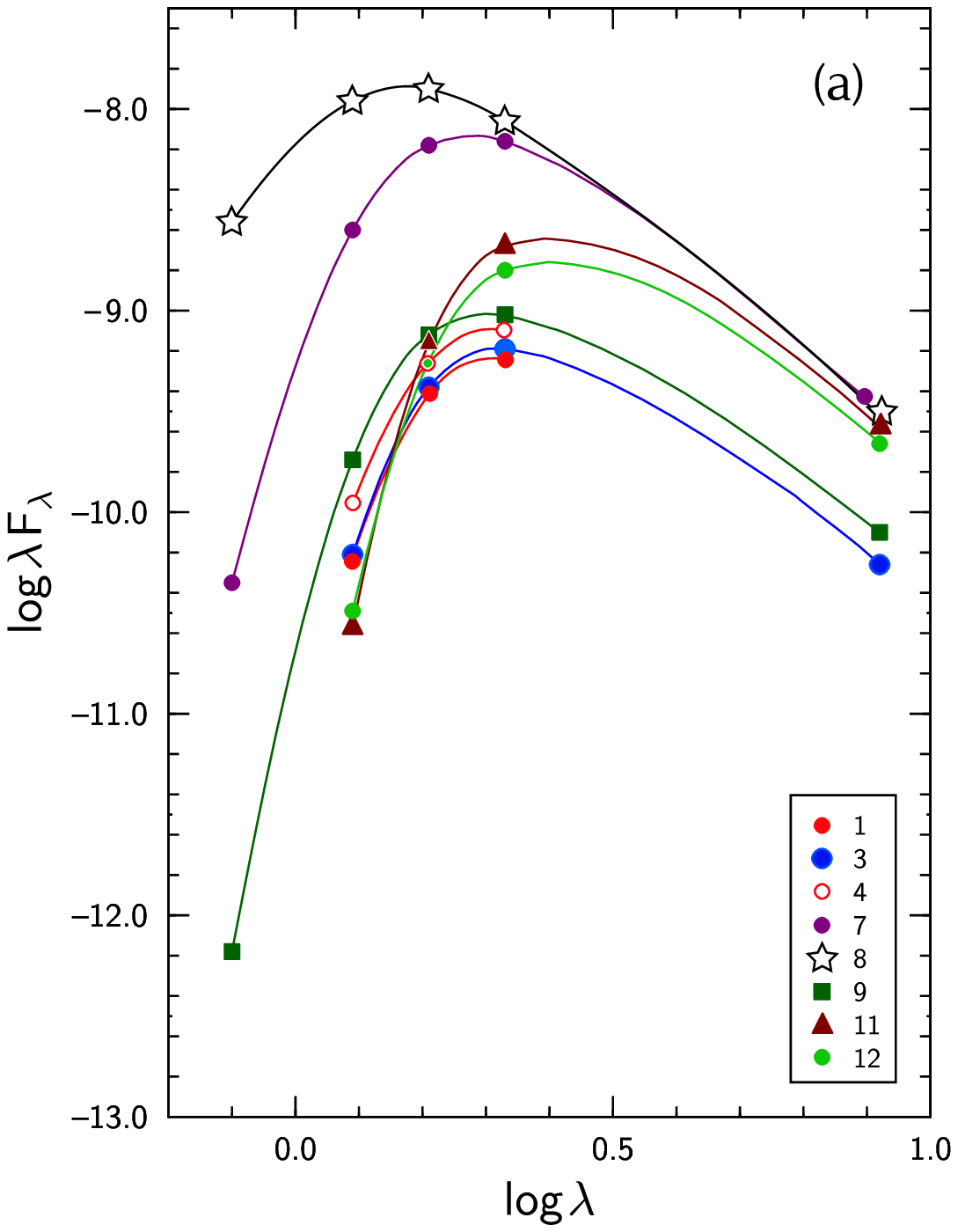,angle=0,width=61truemm,clip=}}
\hskip3mm
\parbox[t]{61mm}{\psfig{figure=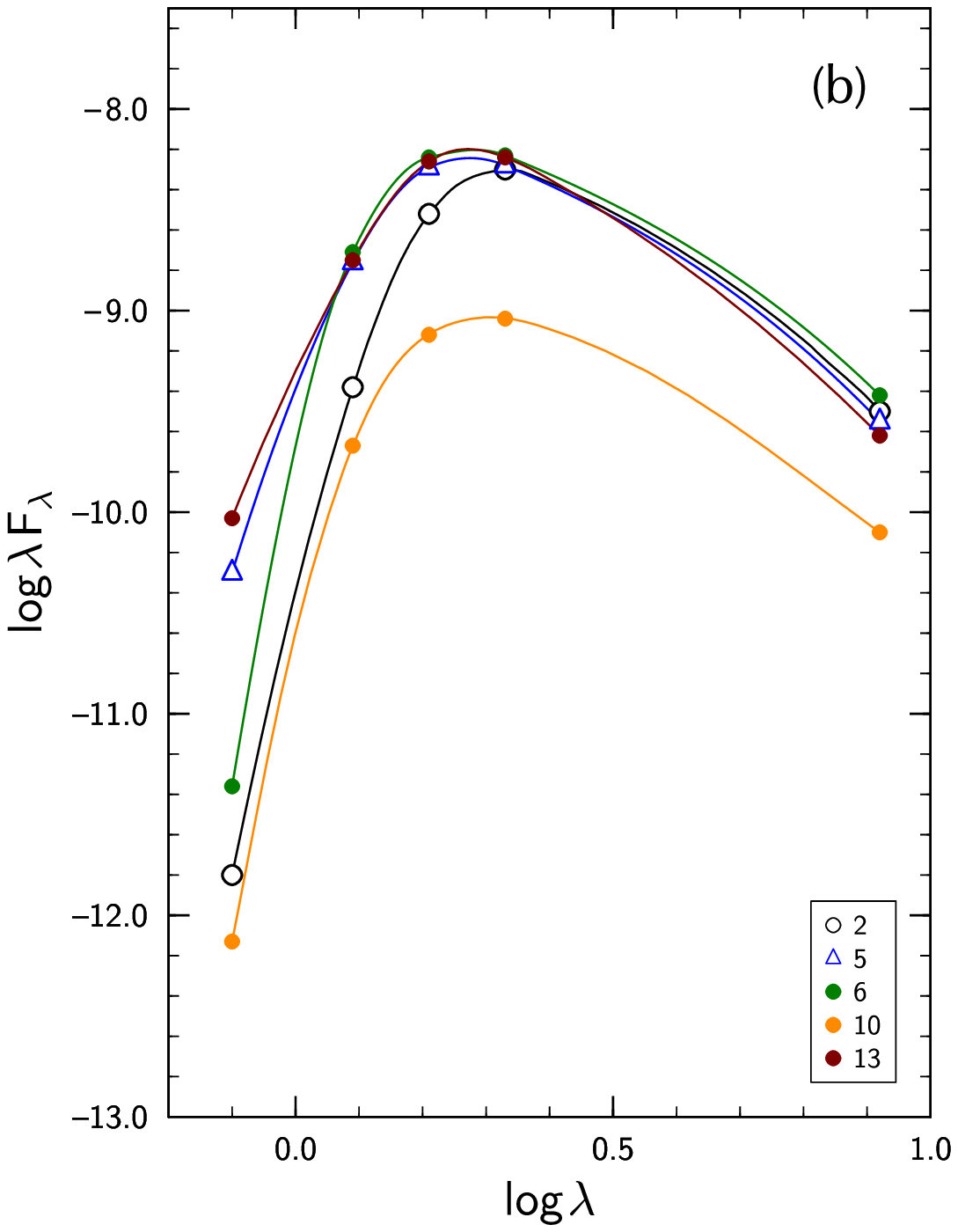,angle=0,width=61truemm,clip=}}
\vskip2mm
\captionb{5}{ Spectral energy distributions of the O-like stars between 0.8
and 8.3 $\mu$m.  Here $\lambda$ is in $\mu$m and $F_{\lambda}$ is in
erg\,$\times$\,cm$^{-2}$\,$\times$\,s$^{-1}$\,$\times$\,$\mu$m$^{-1}$.
Panel (a) shows the stars which are the candidates to O-type stars and
panel (b) the candidates to AGB stars.}
}
\end{figure}


\sectionb{6}{INTERSTELLAR EXTINCTION IN THE DIRECTION OF O-LIKE STARS}

The most detailed and deep investigation of interstellar extinction in
the direction of the NAP complex was published by Cambr\'esy et al.
(2002) who applied the method based on star counts in $K_s$ magnitude
and on statistical color excesses $E_{H-K_s}$.  The angular resolution
of the extinction map is 4--7\arcmin\ in the high extinction areas with
$A_V$ between 20--30 mag.  This means that smaller areas with high
extinction cannot be resolved.


\begin{figure}[!th]
\vbox{
\centerline{\psfig{figure=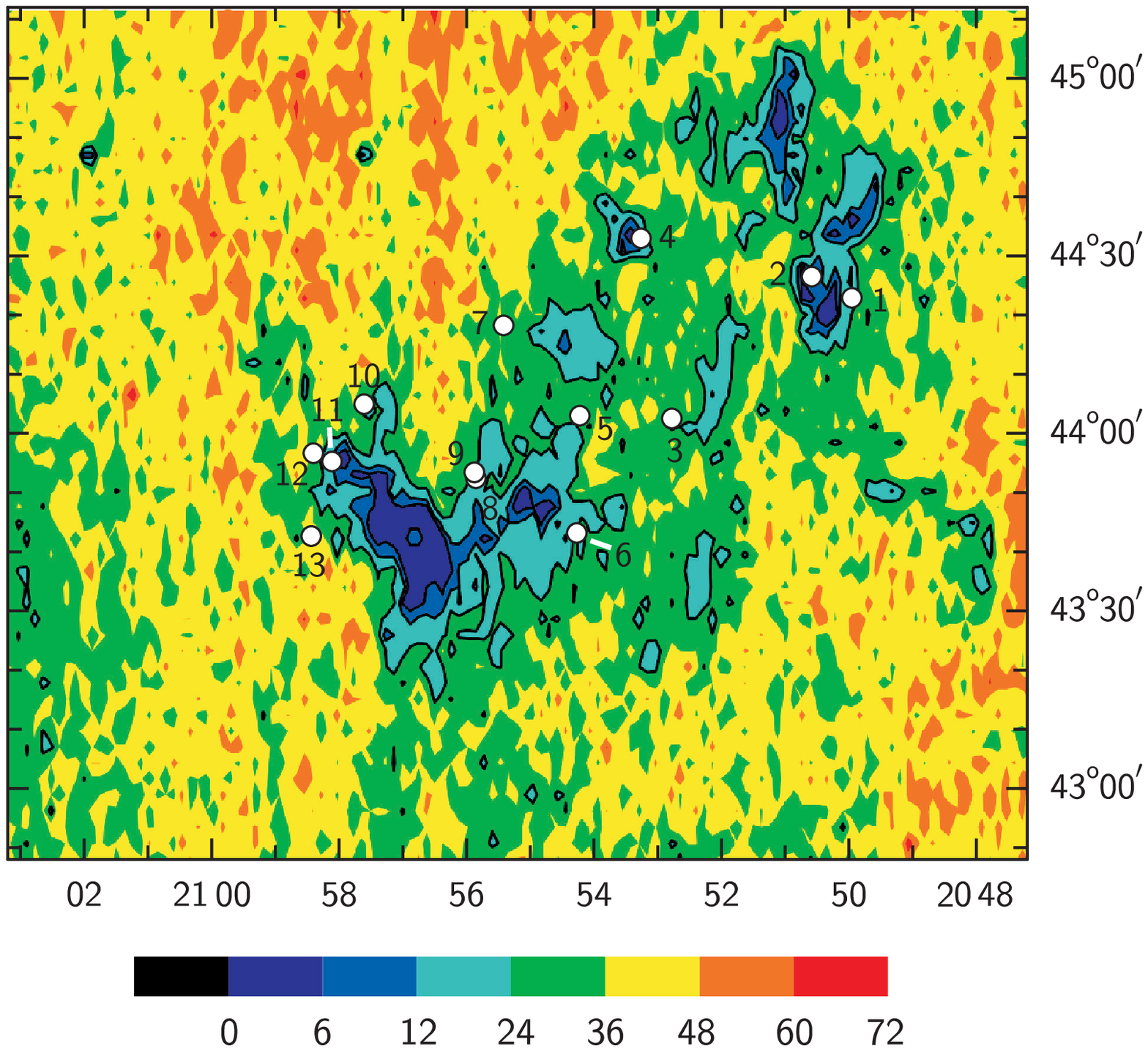,width=124mm,angle=0,clip=}}
\vspace{-.5mm}
\captionb{6}{The map of star counts in the $K_s$ magnitude in
 $2\arcmin \times 2\arcmin$ cells. The scale below the map indicates
numbers of stars. The numbered white circles show the locations
of the O-like stars from Table 1.}
}
\end{figure}

To examine the extinction distribution in greater detail we have applied
star counts in $K_s$ within 2\arcmin\,$\times$\,2\arcmin\ cells.
500\,000 stars with the indication of photometric uncertainty $\sigma <
0.25$ were used.  The number of stars falling in each cell varied from
zero to $\sim$\,70.  The resulting star density map for RA 20$^{\rm
h}$48$^{\rm m}$ -- 21$^{\rm h}$02$^{\rm m}$ and DEC 43\degr\ -- 45\degr\
is shown in Figure 6. It exhibits more details and broader boundaries of
large extinction than the Cambr\'esy et al.  (2002) map.  Since we have
no calibration of star counts in extinctions, the map can be used only
for a qualitative estimate of the extinction in the NAP complex in the
direction of O-like stars listed in Table 1.

We also applied another method to estimate the maximum extinctions in
the vicinities of O-like stars, based on the reddenings of the
background red-clump giants.  From the $J$--$H$ vs.  $H$--$K_s$
diagrams, plotted for the areas of 20\arcmin\ diameters around these
stars, the maximum values of $J$--$H$ were read out taking into account
only stars close to the reddening line with the slope
$E_{J-H}$/$E_{H-K_s}$ = 2.0, originating from the intrinsic position of
red clump giants, i.e., $J$--$H$ = 0.50, $H$--$K_s$ = 0.14 (see
Strai\v{z}ys et al. 2008).  Possible pre-main-sequence stars, identified
by the criterion $Q_{JHK_s}$\,$<$\,0.0, were excluded from
consideration.  Distances to 10 stars with maximum extinctions in each
direction were calculated, accepting their $M_K$ = --1.7 and the
extinctions determined from $E_{J-H}$.  Most of these stars are located
between 0.6 and 2 kpc, i.e., belong to the Local arm.  We accept that
their extinction originates mainly in the NAP complex, since no more
distant dense clouds are known in this direction.  The maximum values of
$A_V$ calculated by the equation:
$$
A_V ({\rm max}) = E_{J-H}~ / ~0.12 = (J-H - 0.50)~ / ~0.12.
$$
are given in Table 1.

\sectionb{7}{COMMENTS ON INDIVIDUAL STARS}
\vskip1.3mm

\noindent {\bf 1 = 2MASS J20495723+4422538}

The star is one of the reddest objects in Table 1. For it only 2MASS
photometry is available.  If the star is of O-type, its extinction $A_V$
must be 22.7 mag.  This is in agreement with the maximum extinction
value of the background red giants given in Table 1. Its proximity
within $\sim$\,10\arcmin\ to the bright radio E-rim (Matthews \& Goss
1980) and to the Pelican Nebula hot spot (Bally \& Scoville 1980) makes
it a good candidate for the ionizing source of the upper part of the
Pelican head.  Its position in the $K_s$ vs.  $H$--$K_s$ diagram (Figure
2) is consistent with what we expect for a star of spectral class close
to B0\,V at a distance of 550 pc.  If the star is more distant, its
spectral type can be earlier.

\vskip1.3mm
\noindent {\bf 2 = 2MASS J20503505+4426296}

Bally \& Scoville (1980) suspected that this star (called as IRS\,4 in
their paper) is responsible for the hot spot of the Pelican Nebula
discovered in CO maps.  However, the $I_{\rm C}$--$J$ vs.  $J$--$H$
diagram (Figure 3) suggests that the star is probably a carbon-rich
object.  In the color-magnitude diagram (Figure 2) it lies about 1~mag
above the reddening line of O5\,V stars.  Presumably, the star can be
matched with the IRAS 20487+4415 source.  Consequently, this star
probably is not related to the Pelican Nebula hot spot.

\vskip1.3mm
\noindent {\bf 3 = 2MASS J20524679+4402308}

Both in the color-magnitude diagram (Figure 2) and in the $J$--$H$ vs.
$H$--$K_s$ diagram (Figure 3) this star lies very close to star No.\,1.
Their spectral energy distributions are also very similar.  On the sky
the star is located in the emission opening below Pelican's beak.  If
the star is of spectral type near O9--B0, its extinction $A_V$ must be
about 23 magnitudes.  This seems to be too high taking into account the
maximum value of $A_V$ = 20.5 given in Table 1 and the absence of a
dense dust condensation in this direction (Figure 6).  Therefore, we
cannot exclude that the star is a heavily reddened carbon- or
oxygen-rich AGB object.  For the verification the far-red magnitude $I$
of the star would be important.

\newpage
\noindent {\bf 4 = 2MASS J20531582+4432570}

This star is located in a dense dust cloud within only a few arcminutes
from the radio-bright J-rim discovered by Matthews \& Goss (1980).  A
bright point-like radio source seen $\sim$\,15\arcmin\ north of the rim,
is not related to the NAP complex -- it is a distant H\,II region in the
Perseus arm (Wendker et al. 1983; Heske \& Wendker 1985).  If the star
is of spectral class O, its $A_V$ must be 21.3 mag, which is in
contradiction to the expected maximum extinction (18.7 mag) given in
Table 1. The latter value of extinction may mean that accidentally no
background red giant is seen in the direction of a small dust
condensation which covers star No.\,4 and gives the extinction close to
21 mag.  In the color-magnitude diagram (Figure 2) the star lies close
to the reddening line of O9--B0\,V type stars.  Consequently, if the
star belongs to the NAP complex at a distance of 500--600 pc, its
ionizing possibility is not great.  However, the star may be responsible
for the creation of the dense ionized rim of the dust cloud.  If the
complex distance is 550 pc, the projected distance between the rim
center and the star is 0.8 pc only.  Its SED (Figure 5a)
is not very informative, since the star has not been observed both in
$I$ and MSX passbands.  However, the SED in the $J$, $H$ and $K_s$
passbands seems to be quite similar to SEDs of stars Nos. 1 and 3.

\vskip1.3mm
\noindent {\bf 5 = 2MASS J20541342+4402589}

The star is located in the dark cloud near the tip of Pelican's beak.
In the color-magnitude diagram (Figure 2) the star lies almost on the
interstellar reddening line of O5\,V stars at a distance of 550 pc.  As
it was shown in Section 4, the star dos not seem to be of spectral class
O; probably it is an asymptotic giant branch object.

\vskip1.3mm
\noindent {\bf 6 = 2MASS J20541626+4343091}

The star is located in the dark cloud below the tip of Pelican's beak.
In the color-magnitude diagram (Figure 2) and the $J$--$H$ vs.
$H$--$K_s$ diagram (Figure 3) the star lies very close to star No.\,5.
The diagram $I_{\rm C}$--$J$ vs.  $J$--$H$ (Figure 4) shows that both of
them are AGB objects, but No.\,6 is much cooler.  This is confirmed also
by their SEDs shown in Figure 5b.

\vskip1.3mm
\noindent {\bf 7 = 2MASS J20552516+4418144}

As it was shown in Section 4, the star satisfies all our criteria for
being an O-type star.  In the color-magnitude diagram (Figure 2) it lies
0.5 mag above the interstellar reddening line of O5\,V stars for a
distance of 550 pc.  This means that the star is either more luminous
than the main-sequence stars or is located closer to the Sun than 550
pc.  If the star is of spectral class O, its $A_V$ should be about 16
mag.  The star is located near the edge of the L\,935 dust cloud at the
North America Nebula coast.  In this direction the expected maximum
extinction is more than sufficient to give $A_V$ = 16.  There is one
more argument in favor of our suggestion that star No.\,7 may be of
spectral class O5:  its SED looks like the additionally reddened SED of
the known O5\,V type star, CP05-4, described in Section 2. Luminosities
of both stars coincide in the MSX passband at 8.3 $\mu$m where the
interstellar extinction is close to zero.

\vskip1.3mm
\noindent {\bf 8 = 2MASS J20555125+4352246}

The Comer\'on and Pasquali star, CP05-4, described in Section 2.

\vskip1.3mm
\noindent {\bf 9 = 2MASS J20555270+4353242}

This star is located very close to CP05-4, in projection they are
separated only by 62\arcsec.  The star lies in the domains of reddened
O-type stars in the three diagrams discussed above:  the color-magnitude
diagram $K_s$ vs.  $H$--$K_s$ (Figure 2) and the two-color diagrams,
$J$--$H$ vs.  $H$--$K_s$ (Figure 3) and $I_{\rm C}$--$J$ vs.  $J$--$H$
(Figure 4).  However, in the color-magnitude diagram it exhibits lower
luminosity than its neighbor, CP05-4; if the star is at 550 pc, its
spectral type must be around O9--B0\,V, and the interstellar extinction
$A_V$ = 19.3 mag.  This value does not rise a problem with the expected
extinction in this direction (Table 1).

\vskip1.3mm
\noindent {\bf 10 = 2MASS J20573647+4404559}

The star is located at the northern coast of the Gulf of Mexico.
Figure 4 shows that the star probably is an AGB star of late M
spectral class.

\vskip1.3mm
\noindent {\bf 11 = 2MASS J20580673+4355141}

The star is located at the left coast of the Gulf of Mexico.  In the
$J$--$H$ vs.  $H$--$K_s$ diagram it lies almost exactly on the reddening
line of O-type stars, at $H$--$K_s$ = 2.09.  Since the star is not seen
in the DSS far red images, it has no $N$ (or $I$) magnitude available.
Consequently, we had no possibility to verify whether it is a non-carbon
star.  In the color-magnitude diagram $K_s$ vs.  $H$--$K_s$ (Figure 2)
the star lies $\sim$\,1 mag above the reddening line of O5\,V stars.
The SED curve of the star in $J$, $H$ and $K_s$ shows a heavy reddening
but in the MSX passband at 8.3 $\mu$m its intensity almost coincides
with the O5\,V star CP05-4.  This means that the star could be of
O5-type with the extinction $A_V$ = 35.2 mag.  Such value of the
extinction is considerably too large for this direction (see Table 1).
However, the star can be heavily obscured by a small dust condensation
which does not contain background red giants which would increase the
value of $A_V$ given in Table 1. Other alternative is to assume that we
have here a very distant AGB star (of M or N type) which is so luminous
that its apparent brightness at 8.3 $\mu$m is equal to that of CP05-4
located at 550 pc.

\vskip1.3mm
\noindent {\bf 12 = 2MASS J20582424+4356386}

The star is only 3.5\arcmin\ from No.\,11 and $\sim$\,1\arcmin\ from the
suspected cluster of infrared sources ([CBJ2002] 3a in Cambr\'esy et al.
2002).  The SEDs of both stars are quite similar (Figure 5a), but
No.\,12 is either slightly warmer or has a lower extinction in its
direction.  Its $H$--$K_s$ = 2.015 and $Q_{JHK_s}$ = --\,0.22, i.e. it
is outside the range accepted for O-type stars.  The star has no excess
at 8.3 $\mu$m (MSX), so it is not YSO.  Probably, this star is a distant
AGB object located far behind the NAP complex.

\vskip1.3mm
\noindent {\bf 13 = 2MASS J20582622+4342385}

The star is located at the southern coast of the Gulf of Mexico.
Although in the $K_s$ vs.  $H$--$K_s$ diagram (Figure 2) the star is
close to the reddening line of O5\,V stars, in the $I_{\rm C}$--$J$ vs.
$J$--$H$ diagram (Figure 4) the star lies among carbon stars.

\sectionb{8}{CONCLUSIONS}

1. In the area of the North America and Pelican nebulae we identified
thirteen stars simulating heavily reddened O-type stars at a distance of
the complex, 550 pc (Table 1).  The stars were selected using the
 $J$--$H$ vs.  $H$--$K_s$ and $K_s$ vs.  $H$--$K_s$ diagrams based on
2MASS data.  One of these stars is CP05-4 classified as O5\,V by
Comer\'on \& Pasquali (2005).  This set of stars may contain O-type
stars, B-type stars of luminosities higher than V, A--F supergiants and
cool AGB stars (both oxygen- and carbon-rich).

2. For eight stars of the set, far-red magnitudes $I$ (including DSS2
$N$ magnitudes) were collected from the literature.  Applying the
$I$--$J$ vs.  $J$--$H$ diagram, two carbon-rich and three oxygen-rich
AGB stars were identified.

3. Spectral energy distributions, based on the {\it I, J, H, K}$_s$ and
MSX photometry, give additional information about the selected stars.

4. To estimate the maximum interstellar extinction in the direction of
the `O-like' stars located behind the dark clouds of the NAP complex, we
used the $J$--$H$ vs.  $H$--$K_s$ diagrams for the supposed background
K-type giants.  The star count map in the $K_s$ passband was also
constructed and used to estimate the interstellar extinction in small
areas of the NAP complex.

5. Considering all the observational data together, we conclude that two
stars in our set, Nos. 1 and 4, possibly are stars of late O subclasses
responsible for the creation of the ionized radio rims E and J
discovered by Matthews \& Goss (1980).

6. Other two stars, Nos. 7 and 9, also have a considerable probability
of being O-type stars.  They both satisfy all photometric criteria for
O-stars at a distance of the NAP complex with the $A_V$ extinctions of
16 and 19 mag.  Star No.\,11 is also a probable O-star of early
subclass.

7. The remaining stars in Table 1 can be heavily reddened cool AGB stars
located at different distances in the background of the NAP complex.
However, we cannot rule out the possibility that some of them still may
be hot stars related to the complex.  Only spectroscopy and photometry
of these stars in the near and middle infrared can give the final
answer.

\thanks{ The use of the 2MASS, MSX, IPHAS, CADC, SkyView, Gator and
Simbad databases is acknowledged.  We are thankful to Edmundas
Mei\v{s}tas and Stanislava Barta\v{s}i\={u}t\.e for their help in
preparing the paper.  Credit for color picture of the North America and
Pelican nebulae:  Adam Block/NOAO/AURA/NSF.}

\enlargethispage{3mm}

\References

\refb Bally J., Scoville N. Z. 1980, ApJ, 239, 121

\refb Bessell M. S., Brett J. M. 1988, PASP, 100, 1134

\refb Cambr\'esy L., Beichman C. A., Jarrett T. H., Cutri R. M. 2002,
AJ, 123, 2559

\refb Carpenter J. M. 2001, AJ, 121, 2851

\refb Comer\'on F., Pasquali A. 2005, A\&A, 430, 541


\refb Dobashi K., Uehara H., Kandori R., Sakurai T., Kaiden M.,
Umemoto T., Sato F. 2005, PASJ, 57, S1

\refb Drew J. E., Greimel R., Irwin M. J. et al. 2005, MNRAS, 362, 753

\refb Droege T. F., Richmond M. W., Sallman M. P., Creager R. P. 2006,
PASP, 118, 1666; CDS catalog II/271

\refb Fiorucci M., Munari U. 2003, A\&A, 401, 781

\refb Fitzpatrick E. L. 1999, PASP, 111, 63

\refb Glass I. S., Schultheis M. 2002, MNRAS, 337, 519

\refb Gonz\'alez-Solares E. A., Walton N. A., Greimel R., Drew J. E.
2008, MNRAS, 388, 89

\refb Groenewegen M.\,A.\,T., Blommaert J.\,A.\,D.\,L. 2005, A\&A,
443, 143

\refb Herbig G. H. 1958, ApJ, 128, 259

\refb Heske A., Wendker H. J. 1985, A\&A, 148, 439

\refb Kaplan D. L., Frail D. A., Gaensler B. M., Gotthelf E. V. et al.
2004, ApJS, 153, 269

\refb Koornneef J. 1983, A\&A, 128, 84

\refb Laugalys V., Strai\v{z}ys V., Vrba F. J., Boyle R. P., Philip
A.\,G.\,D., Kazlauskas A. 2006, Baltic Astronomy, 15, 483

\refb Lynds B. T. 1962, ApJS, 7, 1

\refb Matthews H. E., Baars J.\,W.\,M., Wendker H. J., Goss W. M. 1977,
A\&A, 55, 1

\refb Matthews H. E., Goss W. M. 1980, A\&A, 88, 267

\refb Pottasch S. 1956, BAN, 13, 77


\refb Sharpless S., Osterbrock D. 1952, ApJ, 115, 89

\refb Strai\v{z}ys V. 1992, {\it Multicolor Stellar Photometry}, Pachart
Publishing House,\\ Tucson, Arizona

\refb Strai\v{z}ys V., Corbally C. J., Laugalys V. 2008, Baltic
Astronomy, 17, 125 (this issue)

\refb Strai\v{z}ys V., Laugalys V. 2007, Baltic Astronomy, 16, 327

\refb Strai\v{z}ys V., Lazauskait\.e R. 2008, Baltic Astronomy,
submitted

\refb Taylor A. R., Gibson S. J., Peracaula M., Martin P. G. et al.
2003, AJ, 125, 3145

\refb The P. S., Thomas D., Christensen C. G., Westerlund B. E. 1990,
PASP, 102, 565


\refb Wendker H. J., Benz D., Baars J.\,W\,M. 1983, A\&A, 124, 116

\end{document}